\documentclass[9pt,onecolumn,oneside]{opticajnl}
\journal{opticajournal} % use for journal or Optica Open submissions
\usepackage[version=3]{mhchem}
\usepackage[utf8]{inputenc}
\usepackage{graphicx}
\usepackage{float}
\usepackage{amsmath}
\usepackage[utf8]{inputenc}
\usepackage[english]{babel}
\DeclareUnicodeCharacter{2212}{-}
\DeclareUnicodeCharacter{2191}{\uparrow}
\DeclareUnicodeCharacter{200E}{}
\usepackage{array, makecell}
\usepackage{boldline}
\usepackage{colortbl}
\usepackage{lineno}
\usepackage{xcolor}
\definecolor{urlblue}{RGB}{0, 102, 204}
\usepackage{fix-cm}
\usepackage{ragged2e}
\usepackage{soul}
\usepackage{titlesec}

\newcommand{\supplementarysection}{
  \clearpage
  \titleformat{\section}{\normalfont\Large\bfseries}{S\thesection.}{1em}{}
  \setcounter{section}{0}
  \renewcommand{\thesection}{S\arabic{section}}
}
\setboolean{shortarticle}{true}
\title{Resonance Engineering via Harnessing Anti-Parallel Dipole Image Coupling}
\author{Dip Sarker}
\author{Abdoulaye Ndao$^*$}

\affil{Department of Electrical and Computer Engineering, University of California, San Diego, La Jolla, California 92093, USA}
\affil[]{$^*$a1ndao@ucsd.edu}

\begin{abstract}
Precise control of plasmonic resonances across a broad spectral range is central to the development of tunable optical devices. Yet, achieving both redshifts and blueshifts within a single nanostructure has remained elusive. Here we introduce a metal–dielectric–metal (MDM) nanodisk array that enables bidirectional tuning of resonance wavelengths throughout the near-infrared (NIR) region. The observed spectral evolution follows the plasmon ruler relationship, with unprecedented tuning properties. In particular, we report a record blueshift response of 457.82 nm for a small nanodisk thickness variation of only 5–10 nm, the highest blueshift response demonstrated in plasmonic architectures to date. This platform offers finely tunable resonances spanning an exceptionally wide NIR range, providing new insights into electromagnetic (EM) coupling mechanisms and establishing a foundation for next-generation tunable devices in sensing, optical communications, and dynamic displays.
\end{abstract}

\setboolean{displaycopyright}{false} % Do not include copyright or licensing information in submission.

\begin{document}

\maketitle

\section{Introduction}
In the mid-19th century, Michael Faraday's research marked the birth of plasmonics, an interacting phenomenon between the electric field of EM wave and the collective oscillations of free electrons in metals, by conducting a pioneering and systematic study on the optical properties of gold (Au) leaf, enabling transmission of green and reflection of yellow wavelengths of incident light~\cite{Faraday_1857_RSC}. Since then, it has garnered enormous attention from the scientific community due to its unique properties by engineering light–matter interactions at sub-wavelength scales~\cite{Brongersma_Science_2010, Stockman_JOpt._2018, Bahari_JAP_2016, Park_AP_2025}. By harnessing strong confinement and enhancement of optical fields, plasmonics has enabled diverse applications, including sensing~\cite{Ahmet_NL_2010, Arif_ACS_Nano_2012, Park_Nat._Phy._2020}, enhanced transmission~\cite{Hamidi_OC_2014}, subwavelength imaging~\cite{Pendry_PRL_2000, Liu_Science_2007}, advanced light sources~\cite{Oulton_Nature_2009, Altug_Nat_Phy_2006}, metasurfaces~\cite{Herzberger_AO_1963, Lalanne_OL_1998, Bomzon_PL_2001, Fong_ITAP_2010, Yu_Science_2011, Hsu_OL_2017, Ndao_JOpt_2018, Ha_OEx_2018}, nonlinear optical processes~\cite{Konforty_LSA_2025, Zhou_PRL_2025}, all-optical ultrafast magnetic switching~\cite{Lambert_Science_2017, Waleed_AOM_2024, Waleed_PRB_2024}, optical modulation~\cite{Sounas_Nat_Phot_2017, Haffner_Nature_2018}, cloaking~\cite{Alu_PRE_2005, Pendry_Science_2006}, and integrated photonic circuitry~\cite{Hsu_OEx_2021, Yang_OEx_2021, Yang_OL_2022, Feng_Nat._Pht._2025}. Beyond these demonstrations, the advent of tunable plasmonic devices heralds a new era in optical technology, as the dynamic manipulation of light at the nanoscale is realized through control of geometry, refractive index, local environment, and carrier density. Such precise resonance tuning permits real-time modulation, delivering compact and energy-efficient solutions that augment the performance of applications ranging from sensors~\cite{Hatice_Science_2015, Lal_Nat_Phot_2007} and optical communication~\cite{Boyd_PRL_2005, Lal_Nat_Phot_2007} to dynamic displays~\cite{Duan_Nat_Comm_2017, Frank_Sci_Adv_2020}.

Researchers around the world have explored and investigated numerous plasmonic-based tunable nanostructures by modifying the geometrical and material parameters. Generally, plasmonic resonances in these nanostructures are typically known to exhibit redshift properties, where the resonance wavelength moves toward longer wavelengths, with increasing structural dimensions, while blueshift behavior, where the resonance wavelength moves toward shorter wavelengths, is comparatively rare, challenging, and less explored due to the fixed density of free electrons in noble metals (e.g. Au and Ag) that determine their plasma frequency~\cite{Raza_NanoPht_2013}. However, plasmonic blueshift properties can be achieved from the anti-parallel-mode excitation and strong near-field interactions, which require careful engineering of the material composition, geometry, and coupling conditions. Several experimental studies~\cite{Chen_Plasmonics_2012, Jain_Nano_Lett._2007, Niu_OEx_2012, Asad_ACSAMI_2024, Dmitrii_AOM_2025} showed that the blueshifting behavior of nanoparticle (NP)-based structures originated from the antibonding mode between NPs. For example, an Au/Ag core/shell NP nanostructure exhibited blueshift properties by the generation of a hybridized antibonding mode, when the diameter of the Ag shell increased~\cite{Chen_Plasmonics_2012}. Indeed, initial reports showed that two pairs of Au nanodisks exhibited blueshift resonance using polarization along the interparticle axis~\cite{Jain_Nano_Lett._2007}. However, their designs faced challenges in uniformly distributing the NPs on the substrate and had limited spectral coverage across the visible range and low blueshift response. A subsequent study demonstrated blueshifting behavior in an Au NP–spacer–graphene nanostructure by exploiting anti-parallel dipole image coupling between the nanoparticle and graphene~\cite{Niu_OEx_2012}. Similarly, the nanostructures had the uniformly distributed challenges of NPs on the substrate, limited spectral coverage across the visible range, and low blueshift response with the fragility issue of graphene. More recently, Nauman \textit{et al.} reported a practical realization of dual-directional plasmonic shifts via anisotropic strain redistribution in elastomeric substrates, enabling simultaneous redshift and blueshift responses controlled by mechanical deformation and polarization~\cite{Asad_ACSAMI_2024}. Complementarily, Belogolovskii \textit{et al.} demonstrated a CMOS-compatible approach using visible light trimming of silicon-rich nitride micro-ring resonators, achieving bidirectional refractive index changes that enabled both redshifts of $\sim$49 nm and blueshifts of $\sim$10 nm through controllable thermal annealing mechanisms~\cite{Dmitrii_AOM_2025}. Although their approaches provide valuable experimental insights that complement our design-oriented strategy, highlighting the growing interest in multifunctional plasmonic systems capable of bi-directional spectral tuning, these designs covered only a limited portion of the visible spectrum with low blueshifting properties. Moreover, despite improvements in blueshift response compared to conventional NP-based designs, this improvement is still insufficient for broader practical implementation, such as tunable devices, dynamic displays, and sensors. Additionally, whereas most earlier nanostructures focus on parallel dipole-dipole interactions and hybridized gap modes effects between particles and NPs, the image dipole induced in the bottom Au mirror of anti-parallel dipole image coupling is phase-inverted relative to the nanodisk image dipole, yielding a symmetry-anti-symmetry condition, not present in particles and NPs structures. Furthermore, the bi-directional tuning originated from this mirror-mediated interaction by varying Au nanodisk thickness, and we control the spatial overlap of these anti-parallel dipole image between the underlying Au slab and the Au nanodisk, enabling the highest blue- and red-shift response instead of the unidirectional shifts previously reported in nanoparticle coupling mechanisms. Additionally, these previous studies raise an important question: Can plasmonic nanostructures be engineered to achieve enhanced blueshift response and broader spectral tunability, particularly extending beyond the visible wavelength range to encompass the optical C-band used in telecommunications?

To address this question, we numerically and theoretically investigated a periodic MDM array structure that demonstrated both blue- and redshifted resonance properties for the first time in this letter. The unique blueshifted resonance emerged by exploiting antiparallel dipole-image coupling between the metal nanodisks and the underlying metal slab within the MDM configuration. Through numerical simulations, we explored how the coupling between adjacent dipoles in the metal nanodisks influences resonance.
%%%%%%%%%%%%%%%%%%%%%%%%%%%%%%%
%%%%%%%%%%%%%%%%%%%%%%%%%%%%%%%
\begin{figure}[ht]
\centering
\includegraphics[width=\linewidth]{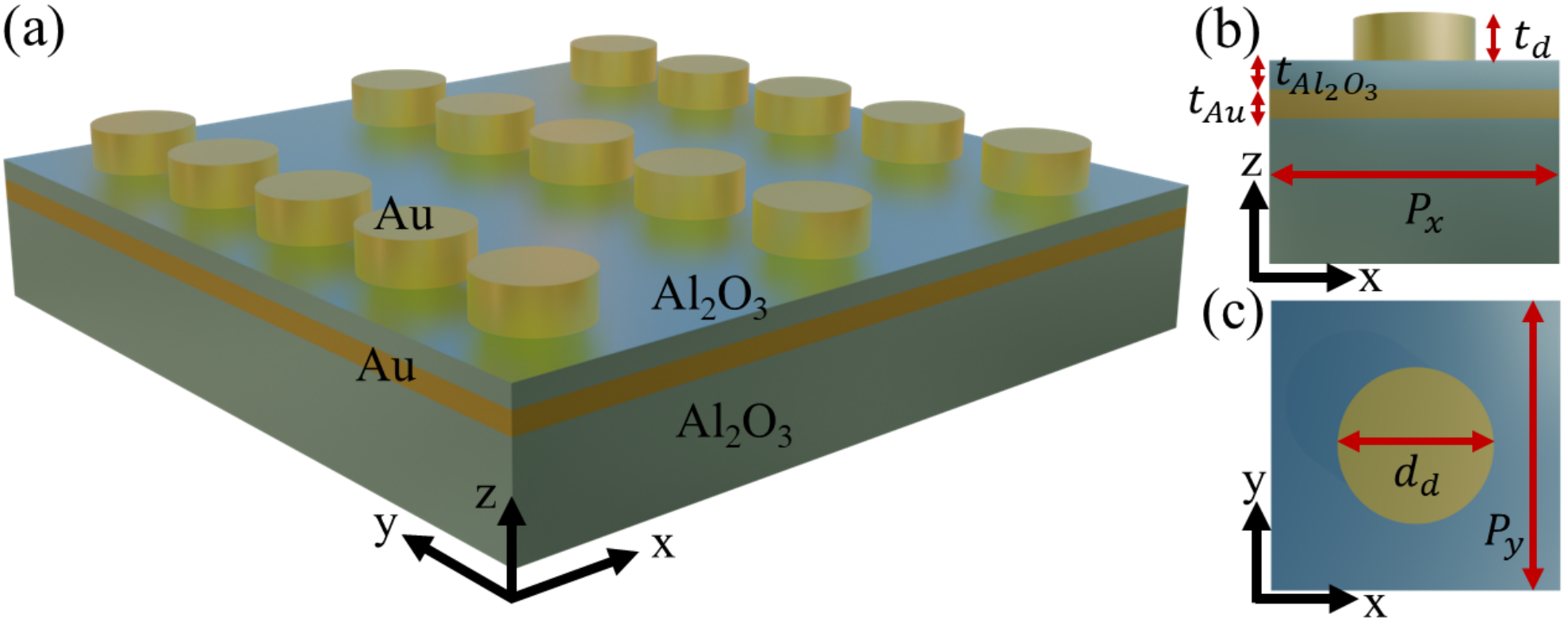}
\caption{\justifying (a) Schematic illustration of the metal–dielectric–metal (MDM) nanostructure for resonance engineering. Au and \ce{Al2O3} were employed as metal and spacer materials, respectively. The nanostructure consists of an \ce{Al2O3} substrate and a top cladding layer of \ce{Al2O3}, enabling the nanostructure to exhibit both blue- and red-shifted resonance behaviors. Cross-sectional (b) xz- and (c) xy-views of the periodic MDM unit cell. Here, $t_d$, $t_{\ce{Al2O3}}$, and $t_{Au}$ represent the thicknesses of the Au nanodisk, \ce{Al2O3} spacer, and Au slab, respectively. The periodicity is denoted as $P_x = P_y = P$, and the diameter of the nanodisk are given by $d_d$.}
\label{Fig. 1}
\end{figure}
%%%%%%%%%%%%%%%%%%%%%%%%%%%%%%%
%%%%%%%%%%%%%%%%%%%%%%%%%%%%%%%
\section{Methodology}
%%%%%%%%%%%%%%%%%%%%%%%%%%%%%%%
To obtain the unique resonance property and understand the physics underlying the phenomenon, the MDM nanostructure on an aluminum oxide (\ce{Al2O3}) substrate was investigated by employing the FDTD method (Ansys Lumerical). Figure~\ref{Fig. 1}(a) presents a schematic of the proposed periodic MDM nanostructure on the \ce{Al2O3} substrate. Au was selected for the nanodisk and bottom slab layers of the MDM plasmonic nanostructure due to its exceptional plasmonic properties, including low optical losses and high chemical stability in the NIR regime. \ce{Al2O3} served as the spacer layer between the nanodisk and bottom slab, facilitating coupling between them. The optical constants ($n$ and $k$) of Au and \ce{Al2O3} were taken from Olmon \textit{et al.}~\cite{Olmon_PRB_2012} and Palik \textit{et al.}~\cite{Palik1998}, respectively. Figures~\ref{Fig. 1}(b) and (c) present the xz- and xy-cross-section views of the unit cell of the nanostructure, respectively. Structural parameters of the nanostructure's unit cell were carefully optimized. The periodicity ($P_x = P_y = P$) of the MDM nanostructure was set as 500 nm. Initially, the Au nanodisk thickness ($t_d$) was selected to be 5 nm, and subsequent variations of $t_d$ were explored to achieve the desired resonance characteristics. To facilitate the anti-parallel dipole coupling between the Au nanodisk and the Au slab, a thickness ($t_{\ce{Al2O3}}$) of 15 nm was employed (see \textcolor{urlblue}{Section 1} of \textcolor{urlblue}{Supplementary 1} for details). The Au slab layer thickness ($t_{Au}$) was chosen as 100 nm to ensure near-perfect reflection and negligible transmission through the nanostructure. $d_d$ and $d_d/2$ denote the diameter and radius of the Au nanodisk, respectively. In our study, $d_d/2$ was set to be 125 nm. To save computational space and time for our numerical study, we used periodic boundary conditions in the x- and y-directions, as our proposed nanostructure was periodic in these directions. In the z-direction, 12 steep angle perfectly matched layers (PMLs) were utilized to absorb the light completely that came out from the simulation region. An override mesh of 1, 1, and 0.4 nm was applied in the x-, y-, and z-directions around the Au nanodisk and spacer layer, respectively. To minimize interference effects, a minimum separation of $\lambda_{center}/4$ was maintained between adjacent objects, where $\lambda_{center}$ denotes the central wavelength of the incident light. We performed our study under the excitation of a plane wave, spanning wavelengths from 800 to 3000 nm. A table of additional simulation informations is presented in \textcolor{urlblue}{Table S1} of \textcolor{urlblue}{Supplementary 1}. To fabricate the proposed MDM nanostructure, we follow the standard cleanroom fabrication recipe, which was developed~\cite{Park_AP_2025, Park_Nat._Phy._2020}. A detailed discussion of the fabrication and the fabrication imperfection analysis and acceptable variation range of the structural parameters is provided in \textcolor{urlblue}{Section 3} of \textcolor{urlblue}{Supplementary 1}.  
%%%%%%%%%%%%%%%%%%%%%%%%%%%%%%%
%%%%%%%%%%%%%%%%%%%%%%%%%%%%%%%
\section{Results and Discussions}
\begin{figure}[ht]
\centering
\includegraphics[width=\linewidth]{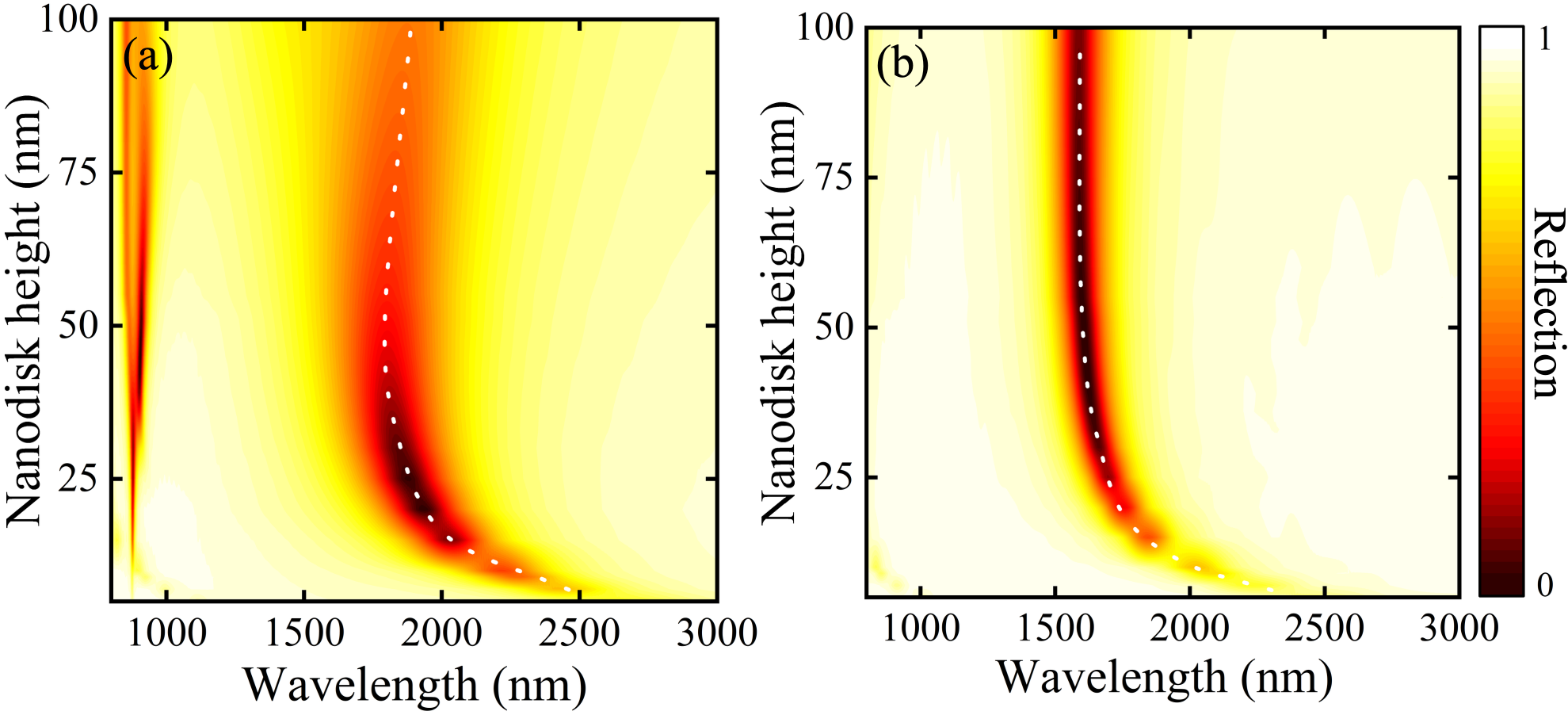}
\caption{\justifying Spatial distributions of the reflection of an MDM nanostructure (a) with and (b) without top cladding layer on the nanostructure.}
\label{Fig. 2}
\end{figure}

%%%%%%%%%%%%%%%%%%%%%%%%%%%%%%%
To analyze plasmonic resonance tuning, we studied our proposed MDM nanostructures by varying the geometrical properties. Specifically, tunable resonances were investigated by changing $t_d$, with and without a \ce{Al2O3} top cladding layer. Figures~\ref{Fig. 2}(a) and (b) show the influence of $t_d$ on the resonance of the reflection spectra with and without the top cladding layer, respectively. In the cladded structure, two resonances appeared in the NIR region. In contrast, the uncladded structure exhibited one resonance in the wavelength range of 800-3000 nm due to the shift of Rayleigh anomaly (RA) towards the visible wavelength.
%%%%%%%%%%%%%%%%%%%%%%%%%%%%%%%
%%%%%%%%%%%%%%%%%%%%%%%%%%%%%%%

\begin{figure}[ht]
\centering
\includegraphics[width=\linewidth]{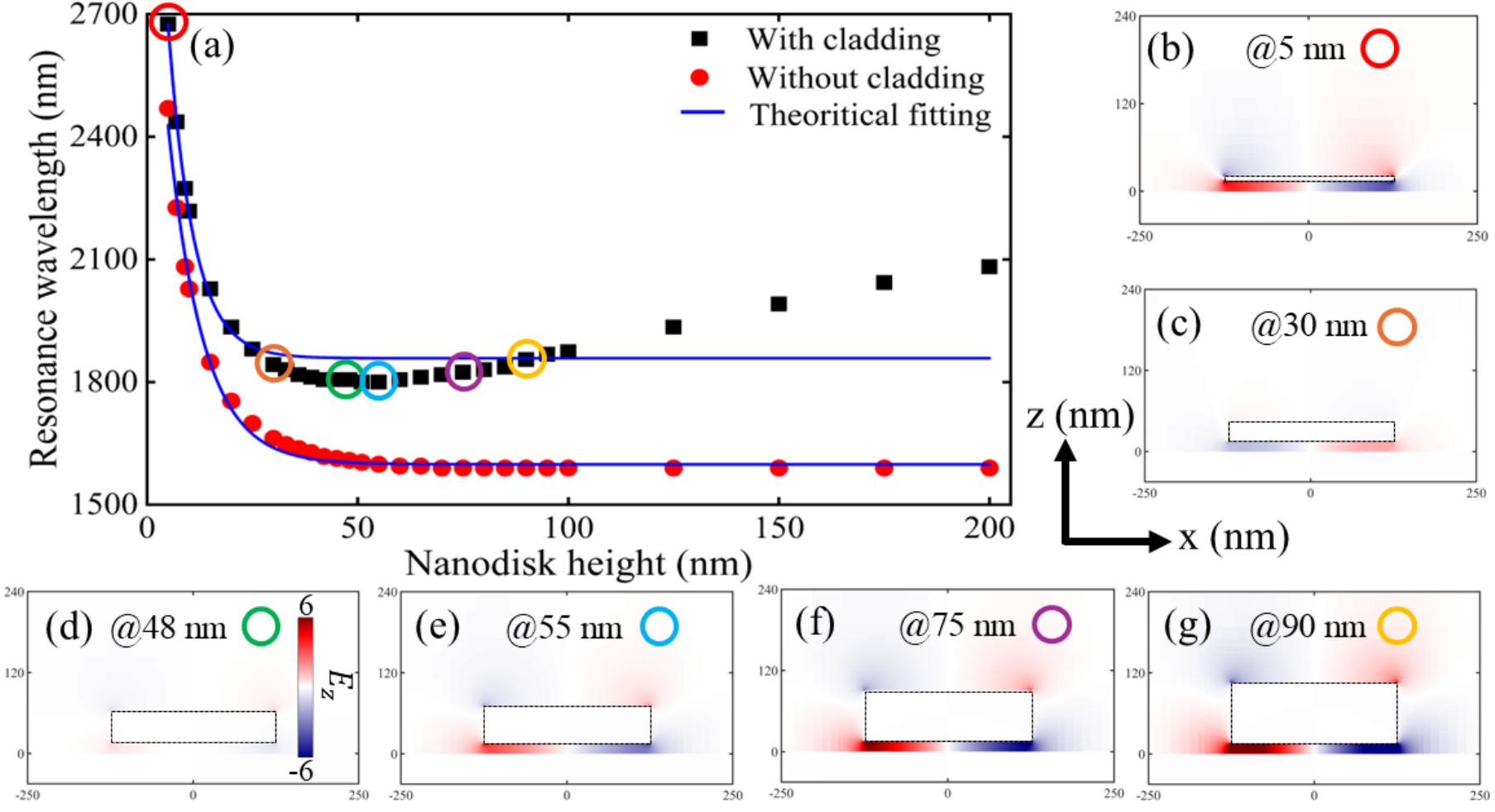}
\caption{\justifying (a) Shift in the plasmon wavelength due to changes in the Au nanodisk thickness with and without a top cladding layer on the nanostructure. The solid lines are theoretical fitting curves by using the plasmon ruler equation. The R$^2$ for these fitted curves was $>$0.9. The xz-view of the spatial electric field ($E_z$) distributions for different Au nanodisk thicknesses (nm) of (b) 5, (c) 30, (d) 48, (e) 55, (f) 75, and (g) 90 for the nanostructure with top cladding layer. The inset of (d) depicts the color bar of $E_z$. The dotted black boxes and rings represent the Au nanodisk and the spatial electric field distributions at different resonance of (a), respectively.}
\label{Fig. 3}
\end{figure}

Figure~\ref{Fig. 3}(a) illustrates the resonance shifts induced by variations in the Au nanodisk thickness, comparing structures with and without a top cladding layer. The longer-wavelength resonance exhibited a blueshift with increasing $t_d$ up to a critical structural parameter. This critical thickness was $\sim$55 nm for the structure with a top cladding layer and $\sim$75 nm for the structure without cladding. Beyond these values, the resonance of the cladded structure redshifted, while it became saturated for the uncladded structure. To intuitively understand this blueshift and redshift behavior with the spatial electric field profiles of the observed blue- and red-shifts, we utilized the anti-parallel dipole image (analogous to an antibonding hybrid plasmonic mode)~\cite{Kik_JPC_C_2010} and perturbation theory~\cite{MTFOL}. We investigated the spatial electric field distributions of the cladded structure' longer-wavelength resonance mode at different $t_d$ (see the circles of different colors in Fig.~\ref{Fig. 3}(a)). When the Au nanodisk is very thin ($t_d<10$ nm), it supports a relatively weak dipole moment. Meanwhile, a thin Au nanodisk supports lower conduction electrons due to its smaller volume which extends its plasmon field in \ce{Al2O3}. Thus, a large fraction of the electric field of the mode is located in the dielectric gap and the surrounding space rather than deep within the Au nanodisk, as shown in Fig.~\ref{Fig. 3}(b). In contrast, when the $t_d$ of the Au nanodisk increases, the Au nanodisk obtains more conducting volume. In the meantime, the increased Au nanodisk achieves more electrons, which results in higher polarizability. Thus, a thicker Au nanodisk confines more of the plasmon’s charge oscillation within the Au nanodisk itself. Therefore, the field penetration into the \ce{Al2O3} spacer layer and the \ce{Al2O3} cladded layer reduces, as shown in Fig.~\ref{Fig. 3}(c). In essence, the image dipole coupling diminishes as the Au nanodisk gets thicker (approaching a critical thickness), as illustrated in Fig.~\ref{Fig. 3}(d). At the critical thickness, the Au nanodisk supports all the dipole moment; thus, the field penetration into the \ce{Al2O3} spacer layer and the \ce{Al2O3} cladded layer is negligible. Therefore, the Au nanodisk behaves almost like it is decoupled from the mirror Au slab, with its plasmon more confined within the Au disk and its immediate vicinity rather than spanning the gap strongly. This behavior mimics the classic plasmon ruler exponential rule: reduced near-field coupling leads to an exponential rise in energy~\cite{Jain_Nano_Lett._2007}. Additionally, near after the critical thickness of $t_d$, the MDM structure becomes from a spacer gap dominated mode (anti-parallel dipole image) to a disk dominated mode. Therefore, the spatial electric field distribution strengthens, as delineated in Fig.~\ref{Fig. 3}(e). As the disk becomes quite thick ($t_d > 55$ nm), it starts to act like a bulkier plasmonic resonator. Moreover, a thicker Au disk has more charge stored for a given field, as shown in Figs.~\ref{Fig. 3}(f) and (g), resulting in redshift. Thus, the redshift property returns. In contrast, this disk-dominated mode cannot exist in the absence of \ce{Al2O3} cladding due to the wavevector mismatch of incident light’s electric field and plasmons of the structure (see \textcolor{urlblue}{Fig. S1} of the \textcolor{urlblue}{Supplementary 1} for details). In addition to anti-parallel dipole image, perturbation theory can explain the effect of metal's small inclusion on resonance. In perturbation theory, the volume perturbation usually tends to redshift the plasmonic mode for thicker Au disk. However, at small $t_d$, the field in the Au volume is relatively trivial, as a large fraction of the electric field of the mode is located in the dielectric gap and the surrounding space rather than deep within the Au nanodisk, as illustrated in Figs.~\ref{Fig. 3}(b) and (c). Therefore, the impact of volume perturbations in the thin-disk regime is initially superseded by the more pronounced influence of image-dipole coupling changes. However, the mode energy increases for large $t_d$ due to the increase of the metal's volume, which results in redshift at the resonance, as shown in Figs.~\ref{Fig. 3}(e)-(g).
%%%%%%%%%%%%%%%%%%%%%%%%%%%%%%%
%%%%%%%%%%%%%%%%%%%%%%%%%%%%%%%

To quantitatively understand this behavior, we analytically studied the longer resonance wavelength ($\lambda_{res,\ 2}$) of our nanostructures as a function of $t_d$ by utilizing the well-known plasmon ruler equation, as shown in Fig.~\ref{Fig. 3}(a). The relation between $\lambda_{res,\ 2}$ and $t_d$ is expressed by~\cite{Jain_Nano_Lett._2007},
%%%%%%%%%%%%%%%%%%%%
\begin{equation}
    \lambda_{res,\ 2} = \lambda_{off} + a e^{-\frac{t_d}{\tau}}.
    \label{Plasmon ruler}
\end{equation}
%%%%%%%%%%%%%%%%%%%%

\begin{figure}[ht]
\centering
\includegraphics[width=\linewidth]{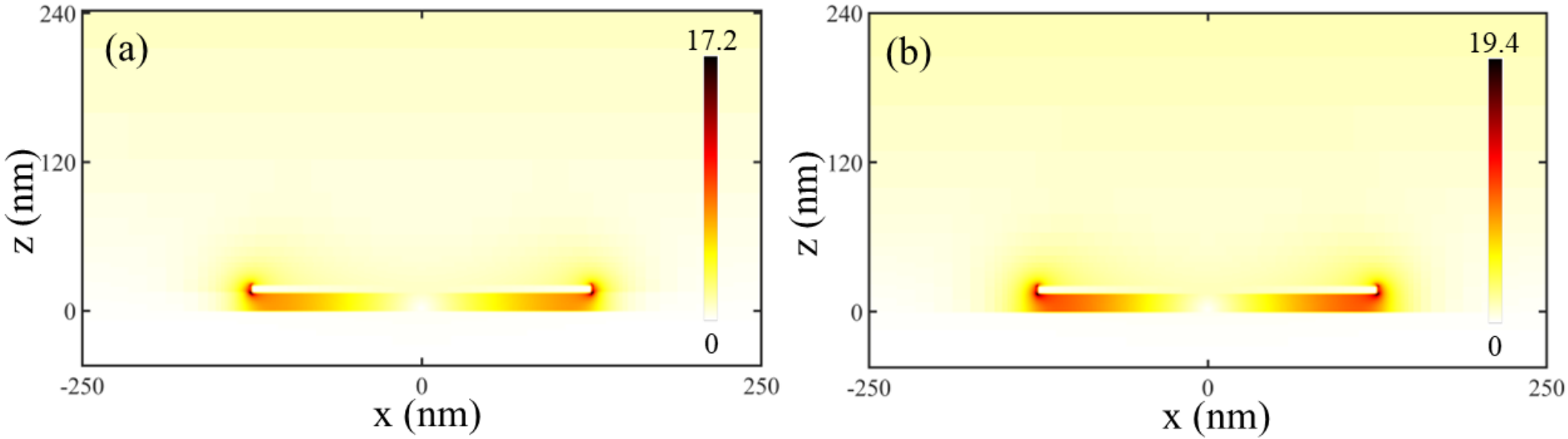}
\caption{\justifying The xz-view of the spatial electric field ($E$) distributions for the nanostructures (a) with and (b) without top \ce{Al2O3} cladding layer, respectively. The color bar is provided in the insets.}
\label{Fig. 4_new}
\end{figure}

The decay length ($\tau$) is defined as the distance at which the coupling decays by a factor of $1/e$. A larger $\tau$ indicates that the plasmon resonance remains sensitive over greater separations, while a smaller $\tau$ confines the strong coupling to shorter distances. $\lambda_{off}$ represents the resonance offset. From theoretical fitting, we obtained $\tau$s of 5.95 and 8.29 nm for the nanostructures with and without a top cladding layer, respectively. The shorter decay length in the presence of a cladding layer leads to greater blueshift response for smaller separations, resulting in enhanced blueshifting behavior (see Fig.~\ref{Fig. 3}(a), $t_d$ range: 5–10 nm). In contrast, the larger $\tau$ for the structure without a cladding layer allows for notable blueshifting over a broader range of $t_d$ (see Fig.~\ref{Fig. 3}(a), $t_d$ range: 35–55 nm), whereas the resonance shift is minimal (blueshift response $\sim$ 18 nm) with the presence of cladding layer. $a$ denotes the strength of the near-field coupling between the Au nanodisk and the underlying Au slab. The values of $a$ were calculated to be 656.72 and 845.05 for the cladded and uncladded nanostructures, respectively, from the theoretical fitting. Consistent with the analytical solution of plasmon ruler equation, numerical simulations demonstrated that the uncladded nanostructure exhibited stronger near-field coupling between the Au nanodisk and Au slab, as shown in Fig.~\ref{Fig. 4_new}. 
%%%%%%%%%%%%%%%%%%%%%%%%%%%%%%%
%%%%%%%%%%%%%%%%%%%%%%%%%%%%%%%

The shorter wavelength resonance was observed only for the cladded nanostructure in the wavelength range of 800-3000 nm. Meanwhile, the shorter wavelength resonance shifted towards the visible wavelength due to the RA phenomenon, specifically influenced by the inclusion of the refractive index of the top cladding layer. Therefore, we need to study the RA phenomenon, explaining the origin of the shorter resonance wavelength ($\lambda_{res,\ 1}$). The relationship among $\lambda_{res,\ 1}$, incidence angle of light ($\theta$), and $P$ is expressed by~\cite{Ndao_JOpt._2014, Su_OEx_2019}, 
%%%%%%%%%%%%%%%%%%%%
\begin{equation}
    \lambda_{res,\ 1}  = \frac{nP \pm sin \theta}{\sqrt{m^2+l^2}}.
    \label{RA}
\end{equation}
%%%%%%%%%%%%%%%%%%%%
%%%%%%%%%%%%%%%%%%%%%%%%%%%%%%%
\begin{figure}[ht]
\centering
\includegraphics[width=\linewidth]{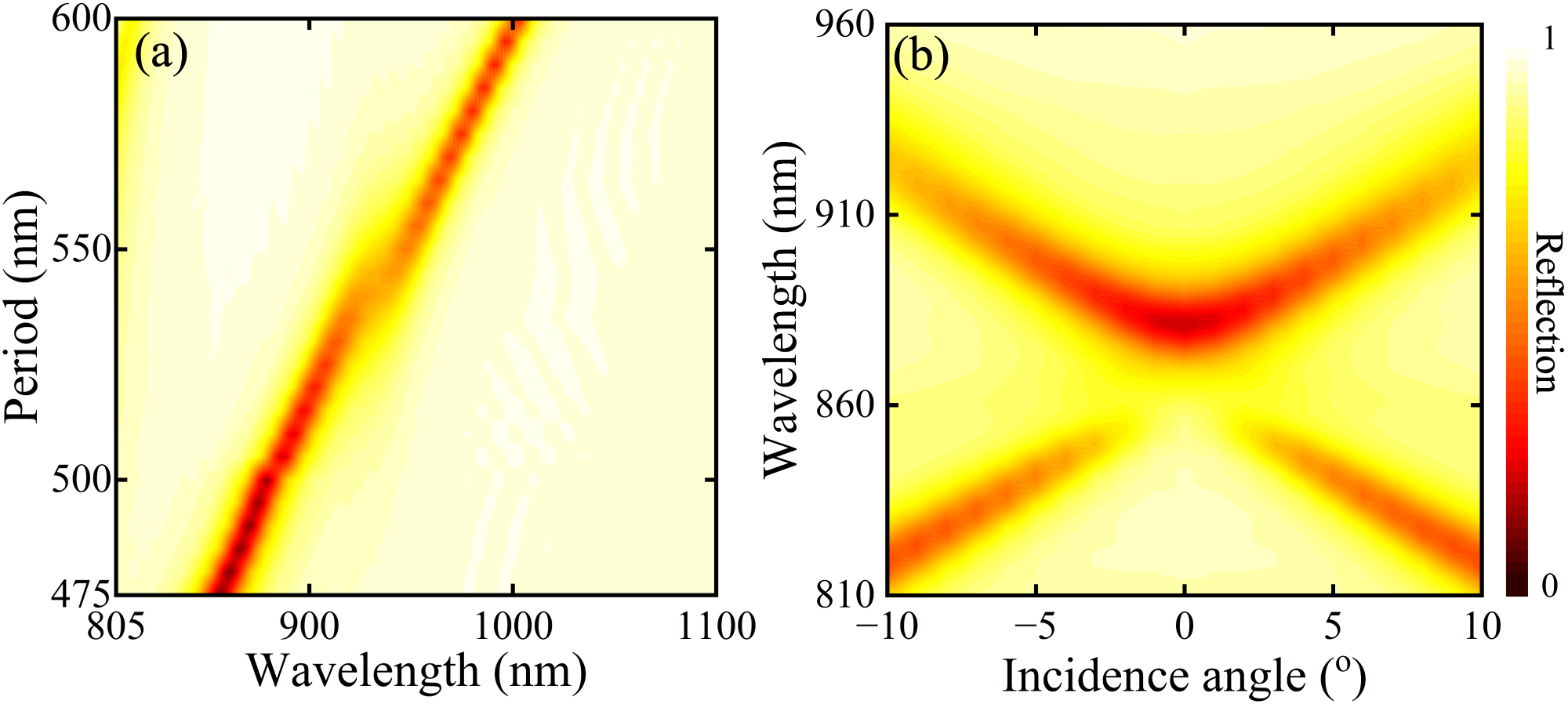}
\caption{\justifying Spatial distributions of the transmission of an MDM nanostructure with top cladding layer for varying (a) $P$ and (b) $\theta$. Here, the $t_d$ and $d_d/2$ were set to be 25 and 125 nm, respectively.}
\label{Fig. 4}
\end{figure}
%%%%%%%%%%%%%%%%%%%%%%%%%%%%%%%

Here, $n$, $m$, and $l$ represent the refractive index of the cladding layer and the diffraction orders, respectively. The ($m$, $l$) was set to be (1, 0) for our study. Between two influencing parameters, we first examined the effect of $P$ on the resonance of the cladded nanostructure with a fixed incident angle of $\theta = 0^\circ$, as shown in Fig.~\ref{Fig. 4}(a). The numerically obtained $\lambda_{res,\ 1}$ of 878.8 nm was in good agreement with the theoretical value of 875 nm for $P = 500$ nm. Second, we analyzed the impact of $\theta$ on $\lambda_{res,,1}$, as depicted in Fig.\ref{Fig. 4}(b), where we set $P$ = 500 nm. A comparison between the theoretical and numerical simulation is provided in \textcolor{urlblue}{Fig. S3} of \textcolor{urlblue}{Supplementary 1}. To investigate the impact of $\theta$, we employed the broadband fixed angle source technique (BFAST), in which a broadband plane wave is incident on the periodic structure at a specific angle. As $\theta$ increased, the resonance split into two branches due to the dependence of $\pm\sin\theta$ in Eq.\ref{RA}. The split is not discernible for small $\theta$; however, the split was observed more than $\theta$ = $1^{\circ}$ in our structure. These results confirmed that the shorter wavelength resonance in the nanostructure arises from the RA phenomenon (see \textcolor{urlblue}{Fig. S4} of the \textcolor{urlblue}{Supplementary 1} for additional information).

\begin{figure}[ht]
\centering
\includegraphics[width=\linewidth]{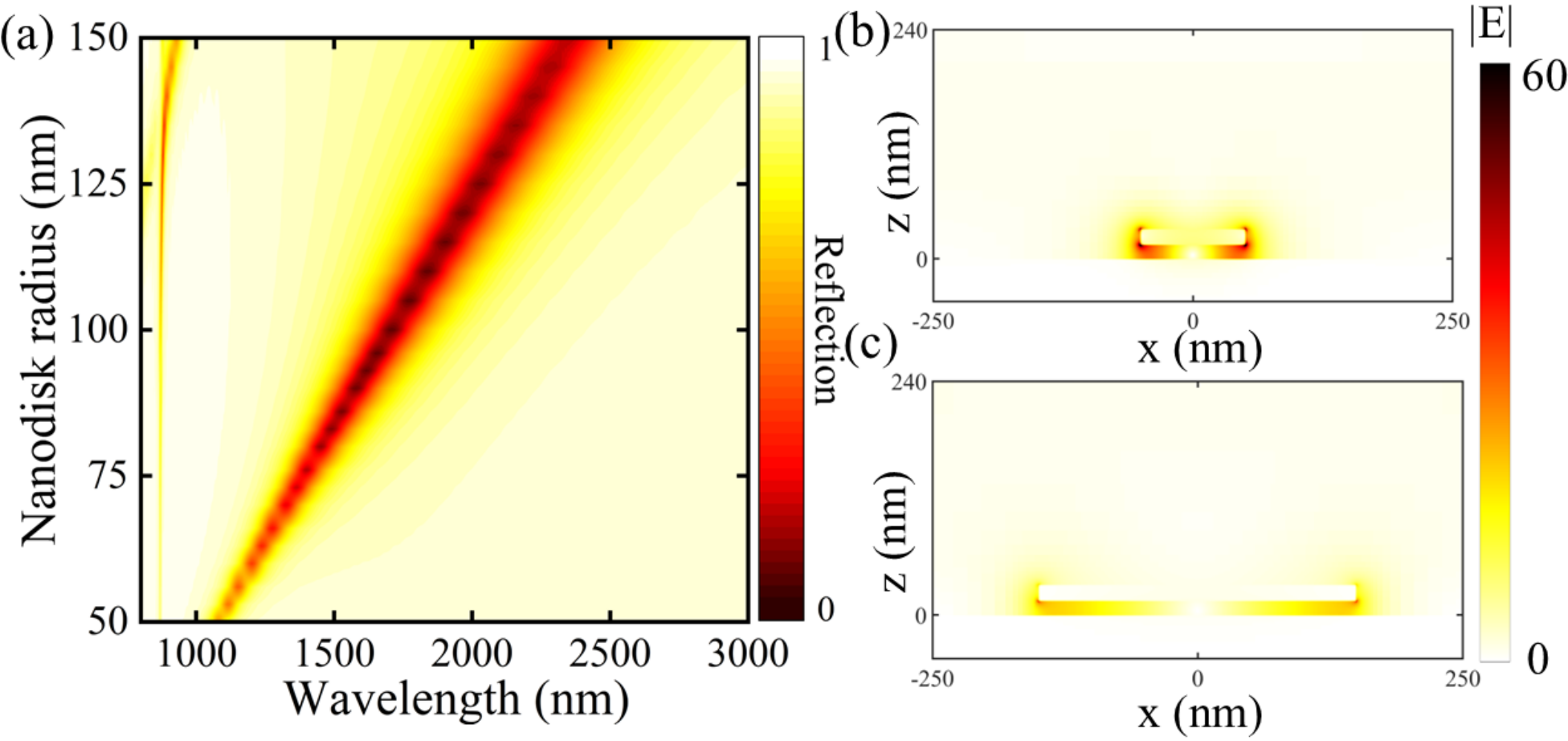}
\caption{\justifying Spatial distributions of the transmission of an MDM nanostructure with a top cladding layer on the nanostructure. We varied the Au nanodisk radius ($d_d/2$) from 50 to 150 nm. Here, the $P$ was set to be 500 nm. The xz-view of the spatial electric field ($|E|$) distributions for the Au nanodisk radius of (b) 50 and (c) 150 nm. The color bar is provided in the inset.}
\label{Fig. 6}
\end{figure}

To examine the impact of Au nanodisk's adjacent dipole coupling on the resonance of the proposed nanostructure, we varied $d_d/2$, where $t_d$ was fixed at 15 nm, keeping all other structural parameters constant. The resonance redshifted with increasing the $d_d/2$ of the Au nanodisk, as shown in Fig.~\ref{Fig. 6}(a). Meanwhile, this redshift resulted from the weakening of nanodisk's adjacent dipole coupling. To visualize this weakening dipole coupling, we investigated the spatial electric field distributions for Au nanodisk radii of 50 nm and 150 nm, as shown in Figs.~\ref{Fig. 6}(b) and (c), respectively. A strong electric field confinement was obtained for the $d_d/2$ of 50 nm. This weakening coupling effect with increasing $d_d/2$ resulted in a wide tunability over a broad NIR wavelength from $\sim$1100 - $\sim$2600 nm, which is not found in previously reported studies to our best knowledge.
%%%%%%%%%%%%%%%%%%%%%%%%%%%%%%%
%%%%%%%%%%%%%%%%%%%%%%%%%%%%%%%

Table~\ref{Table 1} presents a comparative analysis of our proposed nanostructures alongside previously reported state-of-the-art nanostructures for the blueshifting properties. We conducted this analysis, considering the blueshifting response, variation in the structural parameter, normalized bandwidth ($\Delta \lambda_b/ \lambda$), and the operating wavelength. Traditional NP nanostructures, such as Au–Au nanodisk pairs~\cite{Jain_Nano_Lett._2007} and Au–Au nanoparticle pairs~\cite{Rechberger_Opt._Comm._2003}, exhibited blueshifting phenomena primarily within the visible spectral range; however, these structures exhibited limited $\Delta \lambda_b$ for small variation in the structural parameter. Although some improvement was achieved with the introduction of Au–spacer–graphene nanostructure, the $\Delta \lambda_b$, $\Delta \lambda_b/ \lambda$, and tunable range remained insufficient for pragmatic applications~\cite{Niu_OEx_2012}. Notably, current state-of-the-art nanostructures exhibit a higher $\Delta \lambda_b$, up to 79 nm for the structural variation of 9.9 nm, while still operating within the visible wavelength~\cite{Chen_Plasmonics_2012}. However, further development is necessary to extend their operational range and enhance their $\Delta \lambda_b$ sufficiently to enable pragmatic applications. In contrast, our MDM nanostructure demonstrated substantially enhanced $\Delta \lambda_b$, $\Delta \lambda_b/ \lambda$, and wider spectral coverage. Specifically, we achieved a record $\Delta \lambda_b/ \lambda$ of 46.1\% for Au nanodisk thicknesses varying between 5 and 55 nm, a value that is two times magnitude higher than the current state-of-the-art. Even for a small thickness variation from 5 to 10 nm, the $\Delta \lambda_b/ \lambda$ and $\Delta \lambda_b$ remained superior at 24.1\% and 457.82 nm, respectively, compared to the current state-of-the-art. Furthermore, the tunable range of our structure extends deep into the NIR, spanning 1100–2600 nm, which is highly desirable for nanophotonic and sensing applications. This improvement in performance can be attributed to the engineered anti-parallel dipole image coupling, which enables stronger and more localized EM field interactions at subwavelength scales. The exponential decay of near-field coupling, modulated by the nanodisk thickness and spacer properties, provides a powerful means to control the resonance shift with precision. Additionally, the presence of the dielectric cladding layer further enhances the system's responsiveness by shortening the decay length and boosting the field confinement. 
%%%%%%%%%%%%%%%%%%%%%%%%%%%%%%%
\begin{table}[htbp]
\centering
\caption{\bf Comparative analysis}
\begin{tabular}{>{\centering\arraybackslash}p{3.5cm}
                >{\centering\arraybackslash}p{1.5cm}
                >{\centering\arraybackslash}p{3.5cm}
                >{\centering\arraybackslash}p{1.5cm}
                >{\centering\arraybackslash}p{1.5cm}
                >{\centering\arraybackslash}p{2.25cm}}
\hline
Structure & Wavelength & Parameter change (nm) & $\Delta \lambda_b$ (nm) & $\Delta \lambda_b/ \lambda$ (\%) & Ref. \\
\hline
Au-spacer-Graphene & visible & 15 & 29 & 4.5 & \cite{Niu_OEx_2012} \\

Au/Ag NPs & visible & 9.9 & 79 & 16.2 & \cite{Chen_Plasmonics_2012} \\

Au-Au nanodisk pairs & visible & 206 & 150.38 & 23.1 & \cite{Jain_Nano_Lett._2007} \\

Au-Au NP pairs & visible & 250 & 93 & 10.9 & \cite{Rechberger_Opt._Comm._2003} \\

MDM & NIR & 5 & 457.82 & 24.1 & This work \\

MDM & NIR & 50 & 875.78 & 46.1 & This work \\

\hline
\end{tabular}
  \label{Table 1}
\end{table}
%%%%%%%%%%%%%%%%%%%%%%%%%%%%%%%
%%%%%%%%%%%%%%%%%%%%%%%%%%%%%%%
%%%%%%%%%%%%%%%%%%%%%%%%%%%%%%%
\section{Conclusion}
%%%%%%%%%%%%%%%%%%%%%%%%%%%%%%%
In this paper, we demonstrated using MDM array of plasmonic nanostructures that allowed both blueshifts and redshifts in its resonance wavelength in the NIR spectrum. These unique resonance properties are obtained by harnessing antiparallel dipole-image coupling between the metal nanodisks and the underlying metal slab, which follows the plasmon ruler relationship. Most importantly, we obtained a significant blueshift response of 457.82 nm for Au nanodisk thicknesses increasing from 5 to 10 nm. This study demonstrated a single plasmonic platform capable of both blueshifting and redshifting its resonance via simple structural and material modifications. Such dual-shift tunability provides a simple yet powerful means of tailoring resonance wavelengths in the NIR, thereby enabling advanced plasmonic devices and sensors.
%%%%%%%%%%%%%%%%%%%%%%%%%%%%%%%
%%%%%%%%%%%%%%%%%%%%%%%%%%%%%%%
\section*{Disclosures} The authors declare no conflicts of interest.
%%%%%%%%%%%%%%%%%%%%%%%%%%%%%%%
%%%%%%%%%%%%%%%%%%%%%%%%%%%%%%%
%%%%%%%%%%%%%%%%%%%%%%%%%%%%%%%
\section*{Data Availability Statement} 
%%%%%%%%%%%%%%%%%%%%%%%%%%%%%%%
The datasets are not publicly accessible but are available from the corresponding author upon reasonable request.
%%%%%%%%%%%%%%%%%%%%%%%%%%%%%%%
%%%%%%%%%%%%%%%%%%%%%%%%%%%%%%%
\section*{Supplemental Material}
%%%%%%%%%%%%%%%%%%%%%%%%%%%%%%%
See \textcolor{urlblue}{Supplementary 1} for supporting content.
%%%%%%%%%%%%%%%%%%%%%%%%%%%%%%%

% Bibliography
\bibliography{Reference}
\bibliographyfullrefs{Reference}

\supplementarysection

\setcounter{section}{0}
% Reset counters for figures/tables and change their labels
\setcounter{figure}{0}
\setcounter{table}{0}
\titleformat{\section}{\normalfont\Large\bfseries}{\thesection.}{1em}{} 
\renewcommand{\thesection}{S\arabic{section}}
\renewcommand{\thefigure}{S\arabic{figure}}
\renewcommand{\thetable}{S\arabic{table}}
\titleformat{\section}
  {\normalfont\Large\bfseries}   % font
  {S\arabic{section}.}           % label
  {1em}                          % spacing
  {#1}  

\section*{Supplemental Materials}
\addcontentsline{toc}{section}{Supplemental Materials}

\noindent\textbf{\Large Resonance Engineering via Harnessing Anti-Parallel Dipole Image Coupling: Supplemental Material}
%\captionsetup[figure]{labelfont=bf, name=Fig.~, labelsep=period}
%\captionsetup[table]{labelfont=bf, name=Table~S, labelsep=period}

%%%%%%%%%%%%%%%%%%%%%%%%%%%%%%
\section{Impact of Spacer layer thickness on Resonance}
\begin{figure}[htbp]
\centering
\includegraphics[width=\linewidth]{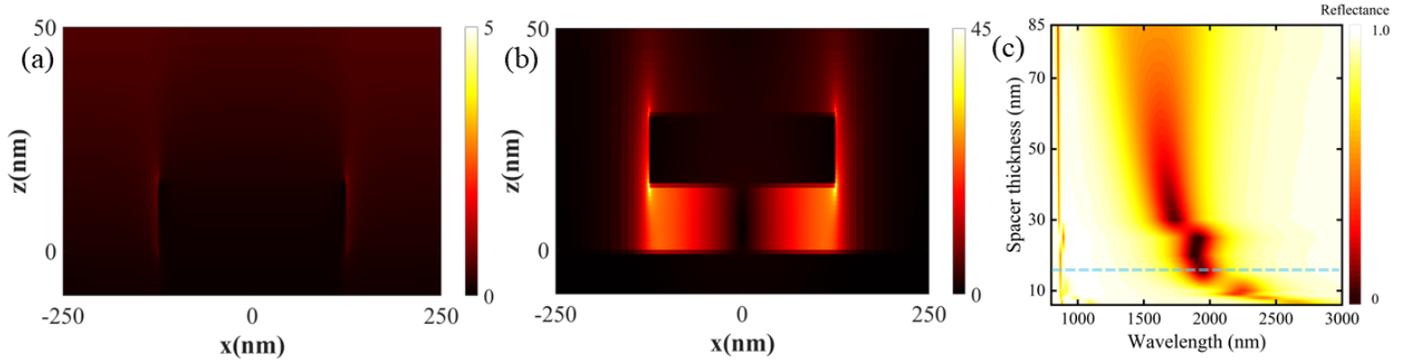}
\caption{The spatial distribution of electric field for (a) $t_{\ce{Al2O3}}$ = 0 nm and (b)  $t_{\ce{Al2O3}}$ = 15 nm. (c) Spatial distributions of the transmission of an MDM nanostructure with a top cladding layer on the nanostructure for varying the spacer thickness. Here, the $t_d$ of 15 nm was considered. The color bar is provided in the inset.}
\label{Fig. R3C1}
\end{figure}

%%%%%%%%%%%%%%%%%%%%%%%%%%%%%%
Figures~\ref{Fig. R3C1}(a) and (b) depict the spatial distributions of electric field for $t_{\ce{Al2O3}}$ = 0 and 15 nm at the longer wavelength resonance. The field confinement of the MDM structure with  $t_{\ce{Al2O3}}$ = 0 nm is negligible compared to that with $t_{\ce{Al2O3}}$ = 15 nm,  indicating that the longer wavelength resonance mode cannot sustain without the spacer layer, as shown in Figs.~\ref{Fig. R3C1}(a). Therefore, it can be inferred that the longer wavelength resonance is governed primarily by the anti-parallel dipole image coupling set by $t_{\ce{Al2O3}}$.

The full width half-maximum (FWHM) and the longer wavelength reflection dip increased with the inclination of spacer layer thickness ($t_{\ce{Al2O3}}$), as depicted in Fig.~\ref{Fig. R3C1}(c). The narrow FWHM and minimum reflection dip provide higher spectral selectivity, lower crosstalk, and higher spectral purity. Therefore, considering the narrow FWHM and the minimum reflection dip, we adopted the $t_{\ce{Al2O3}}$ of 15 nm for our study, as shown in Fig.~\ref{Fig. R3C1}(c) by line cutting.
%%%%%%%%%%%%%%%%%%%%%%%%%%%%%%
%%%%%%%%%%%%%%%%%%%%%%%%%%%%%%
%%%%%%%%%%%%%%%%%%%%%%%%%%%%%%
\section{Simulation Information}
%%%%%%%%%%%%%%%%%%%%%%%%%%%%%%
The simulation information of the FDTD is provided in Table~\ref{Table}.
%%%%%%%%%%%%%%%%%%%%%%%%%%%%%%%
\begin{table}[htbp]
\centering
\caption{\bf FDTD simulation parameters used to investigate the MDM nanostructure}
\begin{tabular}{>{\centering\arraybackslash}p{6cm}
                >{\centering\arraybackslash}p{5cm}}
\hline
Parameter description & Quantities/situation \\
\hline
Simulation type & 3D \\

Plane wave type & Bloch/periodic \\

Spatial cell size (dx = dy = dz) & 10 nm \\

Simulation time & 1000 fs \\

Temperature & 300 K \\

Mesh accuracy & 2 \\

Override mesh size (x, y, and z) & 1, 1, and 0.4 nm \\

Background index & 1 \\

Boundary conditions (x, y, and z) & Periodic, Periodic, and PML \\

Number of PML layers & Steep angle (12) \\
\hline
\end{tabular}
  \label{Table}
\end{table}
%%%%%%%%%%%%%%%%%%%%%%%%%%%%%%%
%%%%%%%%%%%%%%%%%%%%%%%%%%%%%%
\section{Fabrication technique and its imperfection imapct on resonance}
%%%%%%%%%%%%%%%%%%%%%%%%%%%%%%
To fabricate the proposed MDM nanostructure, we will use a standard process that we already developed in our previous work~\cite{Park_Nat._Phy._2020, Park_AP_2025}. The fabrication will be based on three main steps: materials deposition, e-beam lithography, and lift-off. The device will be fabricated on a glass substrate using high-resolution electron-beam lithography (EBL). The substrate is first cleaned by sonication in acetone and isopropyl alcohol (IPA). A chromium adhesion layer is then deposited, followed by a 100-nm-thick gold ground plane and a 100-nm \ce{Al2O3} spacer layer will be deposited using atomic layer deposition (ALD). To minimize sidewall roughness during the lift-off process, a high-resolution positive-tone bilayer resist consisting of methyl methacrylate (MMA EL-8) and polymethyl methacrylate (PMMA A2) is used. After exposure and development in a methyl isobutyl ketone (MIBK) solution, a 3-nm chromium adhesion layer and a gold film are sequentially deposited via electron-beam evaporation. Finally, the resist is removed using a photoresist remover, completing the first device layer.
%%%%%%%%%%%%%%%%%%%%%%%%%%%%%%
\begin{figure}[htbp]
\centering
\includegraphics[width=\linewidth]{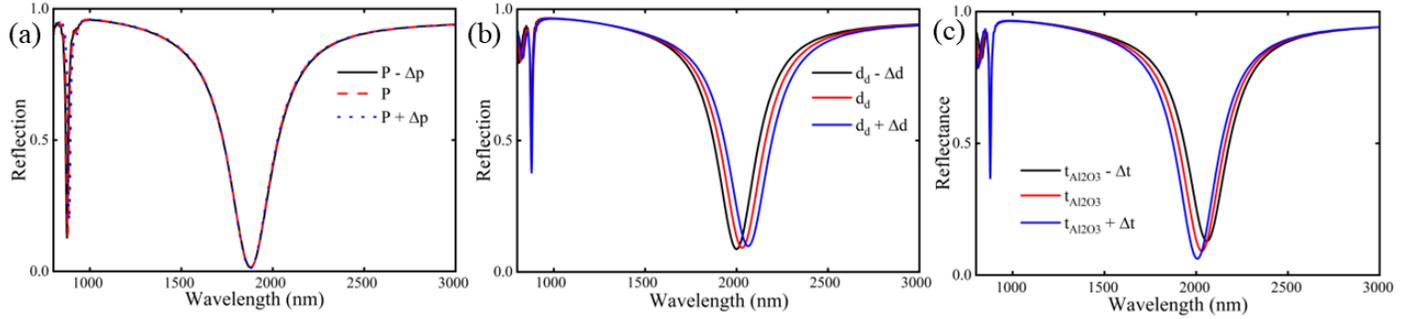}
\caption{Reflection spectra of our MDM nanostructure for imperfection analysis of parameters (a) $P$, (b) $d_d$, and (c) $t_{\ce{Al2O3}}$. Fabrication imperfection analysis variates the dimension at $\Delta p = \Delta d = \pm5$ nm and $ \Delta t = \pm1$ nm.}
\label{Fig. R3C3}
\end{figure}
%%%%%%%%%%%%%%%%%%%%%%%%%%%%%%

To analysis the fabrication imperfection and acceptable variation range of  $P$, $d_d$, and $t_{\ce{Al2O3}}$, we performed a study varying a slight change of  $\Delta p$, $\Delta d$, and $\Delta t$ in these parameters to observe the possible shifts of the resonances in the reflectance spectra, as shown in Fig.~\ref{Fig. R3C3}. In the study, the parameters $\Delta p$ and $\Delta d$ equal to $\pm$5 nm and with the parameter $\Delta t$ equal to $\pm$1 nm were considered. 
As shown in Fig.~\ref{Fig. R3C3}(a), the longer wavelength resonance is insensitive to $P$, and the shorter wavelength resonance exhibits only a negligible shift for $\Delta p$ of $\pm$5 nm. Thus, standard lithographic tolerances in P do not compromise device performance. In contrast, Figs.~\ref{Fig. R3C3}(b) and (c) show that the shorter-wavelength resonance is largely insensitive to $d_d$ and $t_{\ce{Al2O3}}$, whereas the longer wavelength resonance shifts with varying these parameters, as expected. Based on this sensitivity, devices remain within specification for variations up to $\pm 5$ nm in $d_d$ and $\pm 1$ nm in $t_{\ce{Al2O3}}$, with both resonances still clearly resolved. The high resolution of Au nanodisk and thin deposition of the spacer layer can be achievable by employing cleanroom standard lift-off and ALD process.  

Together, from these results, it can be inferred that the shorter wavelength resonance is associated with lattice or diffraction effects set by $P$, whereas the longer wavelength resonance is governed primarily by disk size and anti-parallel dipole image coupling set by $d_d$ and $t_{\ce{Al2O3}}$.
%%%%%%%%%%%%%%%%%%%%%%%%%%%%%%
\section{Rayleigh Anomaly}
\begin{figure}[htbp]
\centering
\includegraphics[width=0.5\linewidth]{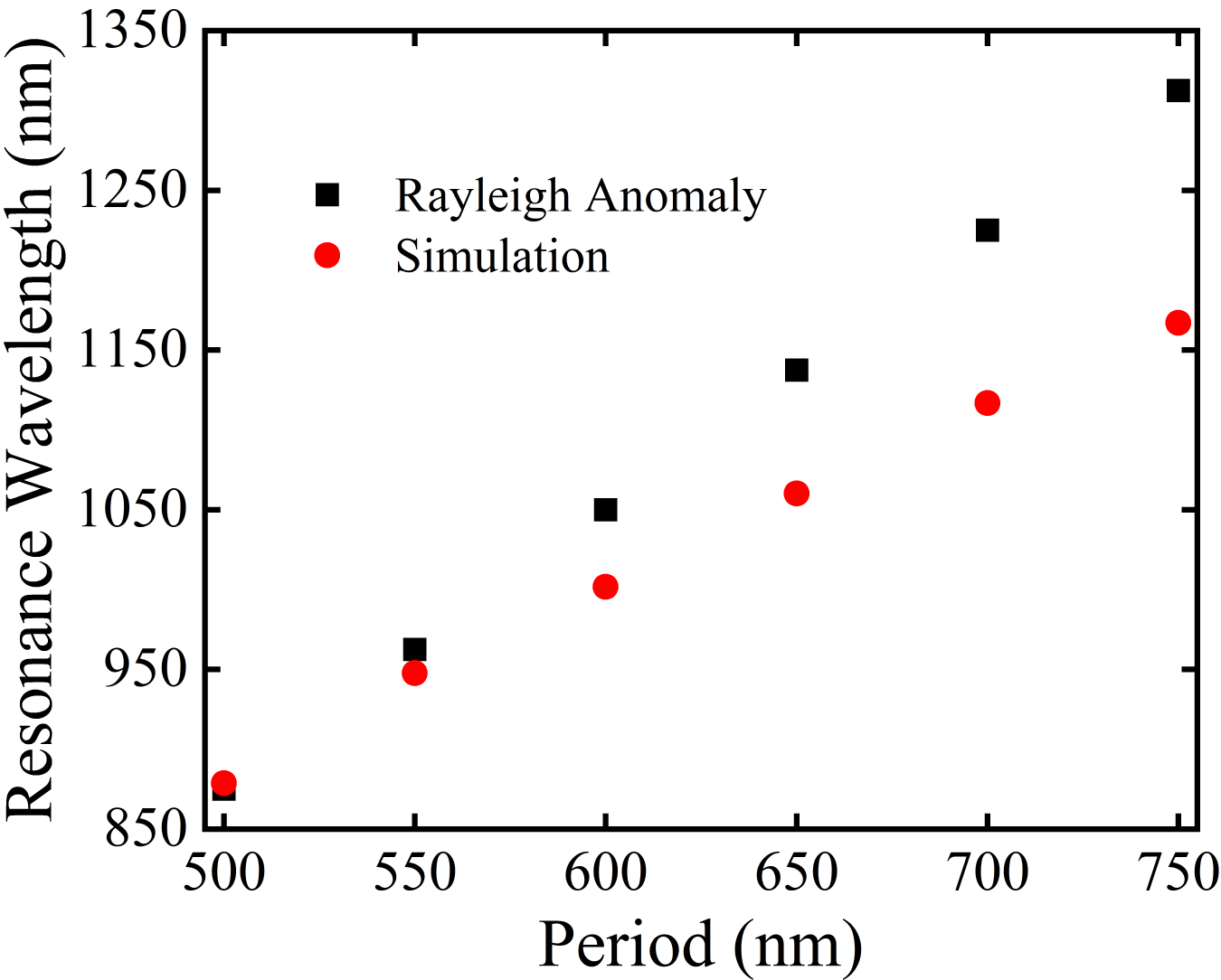}
\caption{Theoretically and numerically obtained resonance wavelength of our MDM nanostructure for different $P$s.}
\label{Fig. S5}
\end{figure}

%%%%%%%%%%%%%%%%%%%%%%%%%%%%%%
We analyzed the impact of the $P$ on the resonance wavelength and compared the resonance wavelength from our theoretical and numerical calculations, as shown in Fig.~\ref{Fig. S5}. The obtained $\lambda_{res,\ 1}$ of 878.779 nm showed a good agreement with the theoretically calculated $\lambda_{res,\ 1}$ of 875 nm for $P$ = 500 nm.   
%%%%%%%%%%%%%%%%%%%%%%%%%%%%%%
%%%%%%%%%%%%%%%%%%%%%%%%%%%%%%   
%%%%%%%%%%%%%%%%%%%%%%%%%%%%%%
\section{Impact of Period and incidence angle}
\begin{figure}[htbp]
\centering
\includegraphics[width=\linewidth]{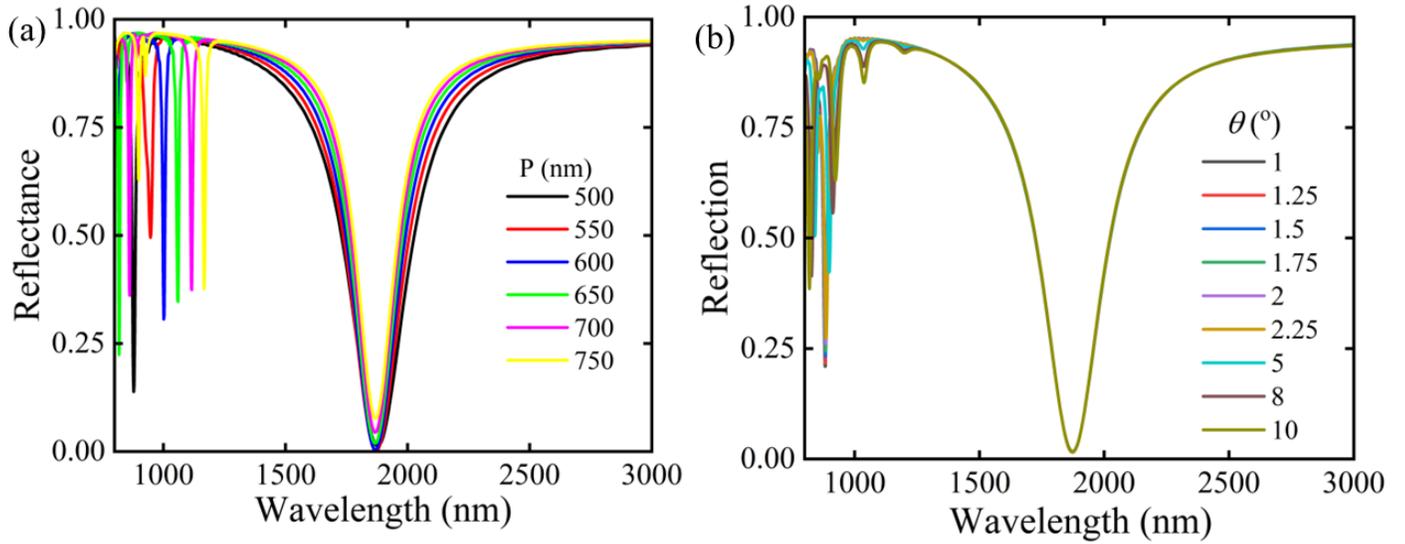}
\caption{The impact of (a) $P$ and (b) $\theta$ on the reflection spectra of our MDM nanostructure.}
\label{Fig. R3C2}
\end{figure}

%%%%%%%%%%%%%%%%%%%%%%%%%%%%%%
We analyzed the impact of the period ($P$) and incidence angle ($\theta$) on the resonance wavelength of our proposed MDM nanostructure, as shown in Fig.~\ref{Fig. R3C2}. As shown in Fig.~\ref{Fig. R3C2}, the longer-wavelength resonance exhibits negligible dependence on $P$ and $\theta$, because of its origin in the localized anti-parallel dipole images. In contrast, the shorter-wavelength resonance arises from the Rayleigh anomaly (RA) and therefore varies with $P$ and $\theta$. As the P increased, the shorter-wavelength resonance $\lambda_{res,\ 1}$ of the MDM array shifted to longer wavelengths, as illustrated in Fig.~\ref{Fig. R3C2}(a). Here, normal incidence was considered for all simulations. For $P$ = 500 nm, The numerically obtained $\lambda_{res,\ 1}$ of 878.8 nm was in good agreement with the theoretical value of 875 nm from RA phenomena. When the $\theta$ is increased, the shorter resonance splits into two branches due to the RA’s $\pm sin\theta$ angular dependence, as shown in Fig.~\ref{Fig. R3C2}(b). These results confirmed that the shorter wavelength resonance in the nanostructure arises from the RA phenomenon.
%%%%%%%%%%%%%%%%%%%%%%%%%%%%%%
%%%%%%%%%%%%%%%%%%%%%%%%%%%%%%

\end{document}